# Upper fields and critical current density of $K_{0.58}Fe_{1.56}Se_2$ single crystals grown by one step technique

Zhaoshun Gao, Yanpeng Qi, Lei Wang, Chao Yao, Dongliang Wang, Xianping Zhang, Yanwei Ma*

Key Laboratory of Applied Superconductivity, Institute of Electrical Engineering, Chinese Academy of Sciences, Beijing 100190, China

**Abstract:**

Single crystals of $K_{0.58}Fe_{1.56}Se_2$ were successfully synthesized by a new single step process with the onset superconducting transition temperature 31.9 K. The x-ray diffraction patterns suggest that they have high crystalline quality and c-axis orientation. A possible modulation structure of Fe-vacancy along c axis was observed. The upper critical field has been determined with the magnetic field along *ab*-plane and *c*-axis, yielding an anisotropy of about 3.3. It has also been shown that the critical current density of the $K_{0.58}Fe_{1.56}Se_2$ is about $1.7 \times 10^4$ A/cm$^2$ at 5 K.

* Author to whom correspondence should be addressed; E-mail: ywma@mail.iee.ac.cn

The discovery of superconductivity in the iron pnictides has triggered great attention in past several years [1-7]. This is the first family with a high transition temperature in a transition metal compound not based on cuprates. Furthermore, these iron based superconductors were reported to have a very high upper critical field, $H_{c2}$, bringing the hope both on more deep understanding of superconducting mechanisms as well as on future applications [4–10]. Up to now, several groups of iron based superconductors have been discovered, commonly denoted as LnOFeAs (1111 series) [1-3], BaFe$_2$As$_2$ (122 series) [5], LiFeAs (111 series) [11-12] and FeSe (11 series) [6]. Very recently, by intercalating alkali metals or Tl into between the FeSe layers, superconductivity around 30 K has been achieved [13-17]. The transport properties of these new superconductors are highly sensitive to its stoichiometry or Fe vacancy. For example, the K$_x$Fe$_2$Se$_2$ could be tuned from insulating to superconducting state by varying the ratio of starting materials or post-annealing [15, 18, 19]. In addition, the FeSe112 series have relatively high $T_c$ of 32 K and no toxic arsenic. Therefore, the FeSe based materials deserve intensive studies for fundamental physics, material growth process and potential applications.

The common procedure for producing FeSe 122 single crystals is two-step sealed quartz tube method [13, 20]. Due to the high activity of alkali metals, the quartz tube would be explosion during the growth procedure. Furthermore, using FeSe as a starting material, it is very difficult to control the Fe content or vacancy. In order to simplify the fabrication process, avoid the explosion accident and to control the stoichiometry more precisely, here we report the preparation of K$_{0.58}$Fe$_{1.56}$Se$_2$ single crystals using our developed single step sintering process.

The single crystals of K$_{0.58}$Fe$_{1.56}$Se$_2$ were grown in the following way. The starting materials of Fe, and Se were mixed in appropriate stoichiometry. Then proper amount of K pieces and Fe + Se mixture were placed in an alumina oxide crucible, and sealed in an arc-welded pure iron tube. The sample was put into tube furnace and heated to 1050 $^o$C slowly and held there for 2 hours, and then was cooled to 780 $^0$C with a rate of –6 ºC/h to grow the single crystals. The obtained single crystals show the dark shiny surface and are easily cleaved into plates, as shown in inset of Fig. 1. The crystals are stable in air and the degradation of superconductivity was not observed after being allowed to keep at air atmosphere for more than one month.

The phase identification and crystal structure investigation were carried out using x-ray diffraction (XRD). Standard four probe resistance and magnetic measurements were carried out using a physical property measurement system (PPMS).

Fig. 1 shows the X-ray diffraction pattern of $K_{0.58}Fe_{1.56}Se_2$ single crystal. Only sharp *00l* (*l* = 2n) reflections were recognized in Fig. 1, indicating that the single crystal was perfectly oriented along c-axis. It is worthy to note that there is another series of (*00l*) peaks (marked by the asterisks), implying that there may be a modulation structure along c axis due to the existence of Fe-vacancy. This result was also observed in $Tl_{0.58}Rb_{0.42}Fe_{1.72}Se_2$ single crystals [21].We should note that the $Rb_2Se$, $Tl_2Se$, Fe and Se powders were used as starting materials in Ref [21]. It indicates that the formation of this modulation structure is easier by using Fe and Se elements as the starting materials than FeSe. The actual composition of the crystals was estimated to be $K_{0.58}Fe_{1.56}Se_2$ by an average of four different points EDX measurements.

Fig. 2 displays the temperature dependence of the normalized resistance for a single crystal of $K_{0.58}Fe_{1.56}Se_2$. From this figure, a sharp drop was observed below the onset temperature of 31.9 K. The normal state resistance of this sample shows a broad bump around 230 K and exhibits metallic behavior below 230 K. The bump anomaly may be a metal-insulator transition. The diamagnetic transition of the sample is shown in the inset of Fig. 2. The ZFC curve shows perfect diamagnetism in the low temperature region and sharp transition.

The temperature dependence of resistivity with various magnetic fields applied along *ab*-plane and *c*-axis were presented in Fig. 3 (a) and (b). The upper critical fields ($H_{c2}$) of single crystal $K_{0.58}Fe_{1.56}Se_2$ were determined using a criterion of 90% points on the resistive transition curves. The estimated upper critical fields were plotted in Fig. 4 as a function of temperature. It is clear that the curves of $H_{c2}$ (T) are very steep with the slopes of $-dH_{c2}/dT|_{Tc}$ = 12.5 T/K for *H//ab* and $-dH^c_{c2}/dT|_{Tc}$ = 3.8

T/K for *H//c*. According to the Werthamer-Helfand-Hohenberg formula [22], $H_{c2}(0) = 0.693 \times (dH_{c2}/dT) \times T_c$. Taking $T_c$= 31.9 K, we can get the values of upper critical fields close to zero temperature limit: $H^{ab}_{c2}$= 276 T and $H^{c}_{c2}$= 84 T. The anisotropy parameter of $H^{ab}_{c2}/H^{c}_{c2}$ is about 3.3 which is consistent with previous reports [18, 23]. These high values of $H_{c2}$ and low anisotropy indicate that this new superconductor has an encouraging application in high fields.

Magnetization hysteresis loops for *H//c*-axis were measured at different temperatures as shown in Fig. 5. We calculated the $J_c$ based on the Bean critical state model $J_c = 20\Delta M/a(1-a/3b)$, where $\Delta M$ is the height of the magnetization loop, and *a* and *b* are the dimensions of the sample perpendicular to the magnetic field, *a < b*. Fig. 6 shows the field dependence of $J_c$. The calculated $J_c$ at 5 K reaches $1.7 \times 10^4$ A/cm$^2$ at 0 T, more than one order of magnitude compared to the result was reported by P. C. Canfield's group [24], which has a $J_c$ value about $10^3$ A/cm$^2$. However, the $J_c$ value at 5 K obtained for $K_{0.58}Fe_{1.56}Se_2$ single crystal is significantly lower than that seen in other FeAs superconductors which generally attained $10^6$ A/cm$^2$ at 5 K [25]. As we known, the electrical transport properties of this new superconductor is very sensitive to its stoichiometry or Fe vacancy [15, 18, 19]. The low $J_c$ values may be ascribed to the local inhomogeneous of the stoichiometry or disordered Fe vacancy. With the passage of time, a higher $J_c$ could be expected in the future for FeSe 122 superconductors with improved fabrication process.

In conclusion, we have developed a safe and simple technique for growing $K_{0.58}Fe_{1.56}Se_2$ single crystals. The magnetic susceptibility and resistivity measurements exhibit sharp superconducting transition indicating good superconducting properties of our sample. A high upper critical field $H_{c2}^{ab}(0)$ of 276 T was obtained. The anisotropy of the superconductor determined by the ratio of $H^{ab}_{c2}/H^{c}_{c2}$ is about 3.3. The $J_c$ value of current sample is still very low and has much potential to be improved by optimization of fabrication method.

**Acknowledgement**

The authors thank Profs. Haihu Wen, Liye Xiao and Liangzhen Lin for their help and useful discussions. This work is partially supported by the National '973' Program (Grant No. 2011CBA00105) and National Natural Science Foundation of China (Grant No. 51002150 and 51025726).

# Captions

Figure 1   The single crystal x-ray diffraction pattern of $K_{0.58}Fe_{1.56}Se_2$. Inset shows the photography of the single crystal (length scale 1 mm).

Figure 2   Temperature dependence of resistivity for $K_{0.58}Fe_{1.56}Se_2$ single crystal at zero field up to 300 K. The inset shows the temperature dependence of dc magnetization for ZFC and FC processes at a magnetic field of H = 20 Oe.

Figure 3   (a) and (b) show the temperature dependence of resistivity for $K_{0.58}Fe_{1.56}Se_2$ single crystal with the magnetic field parallel to the ab-plane and c-axis up to 9 T with an increment of 1 T.

Figure 4   The upper critical fields of $K_{0.58}Fe_{1.56}Se_2$ single crystal for magnetic fields parallel to the ab-plane and c-axis respectively.

Figure 5   Magnetic hysteresis loops of $K_{0.58}Fe_{1.56}Se_2$ single crystal measured at 5, 10, 20, and 25 K.

Figure 6   Magnetic field dependence of $J_c$ at 5, 10, 20, and 25 K for $K_{0.58}Fe_{1.56}Se_2$ single crystal.

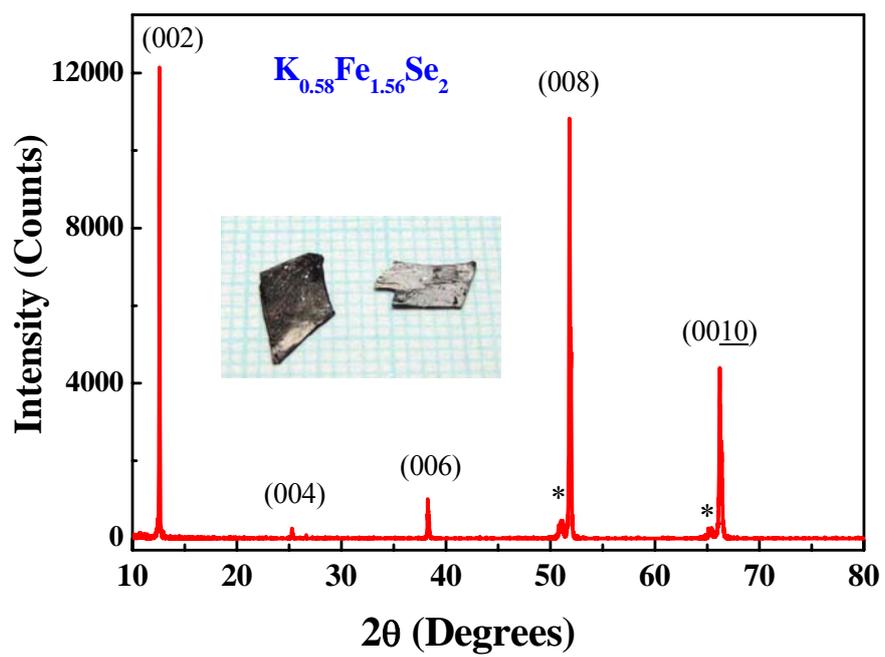

Fig.1 Gao et al.

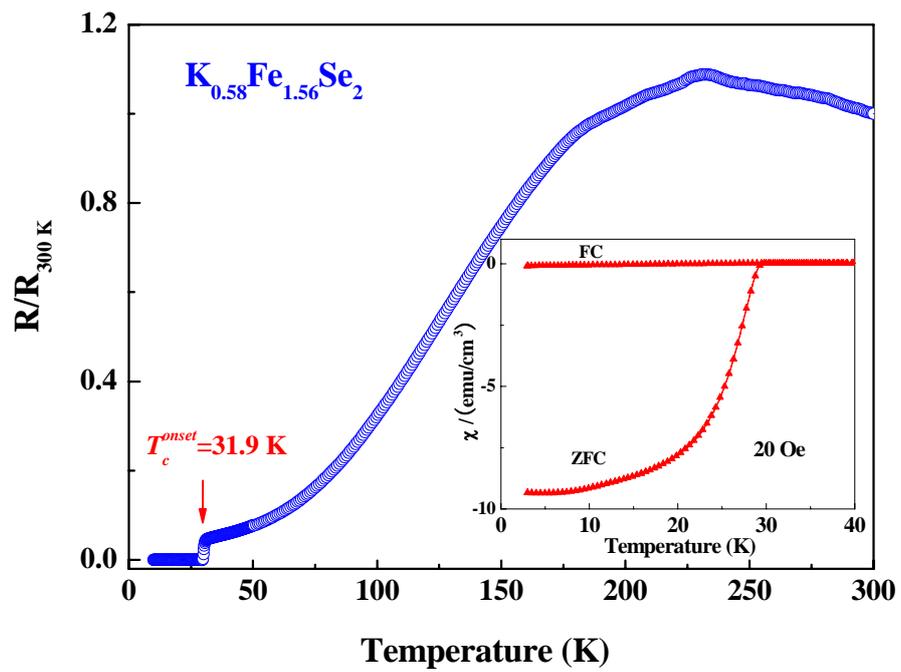

Fig.2 Gao et al.

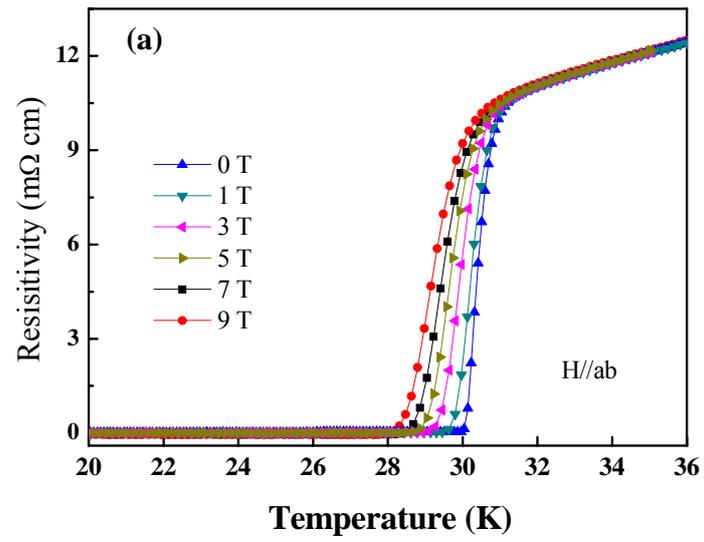

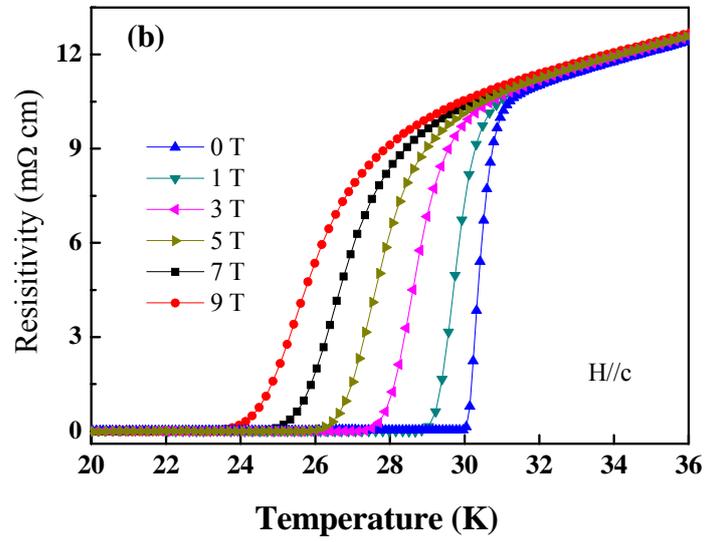

Fig.3 Gao et al.

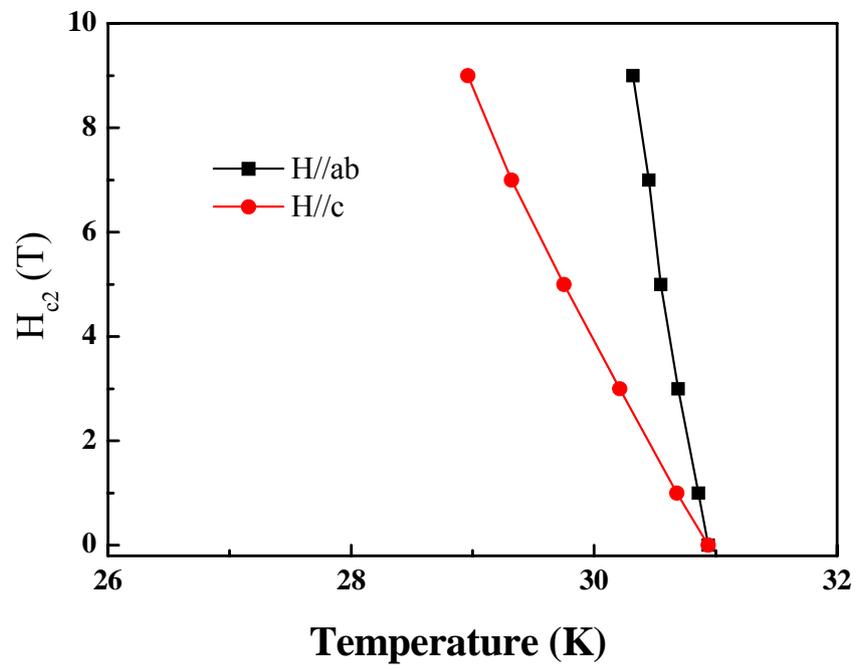

Fig.4 Gao et al.

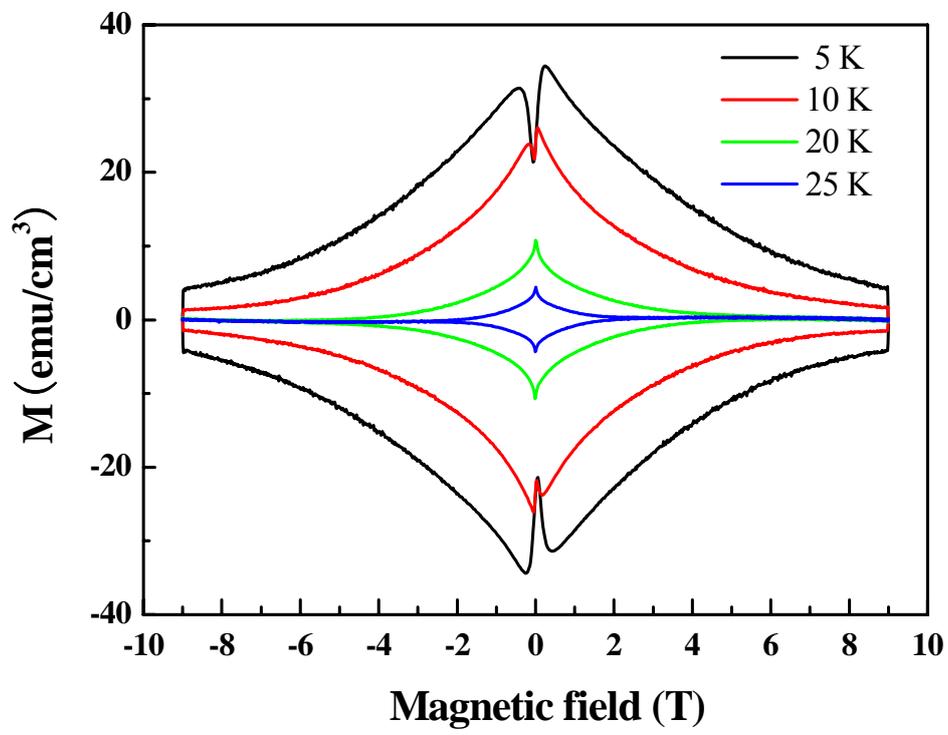

Fig.5 Gao et al.

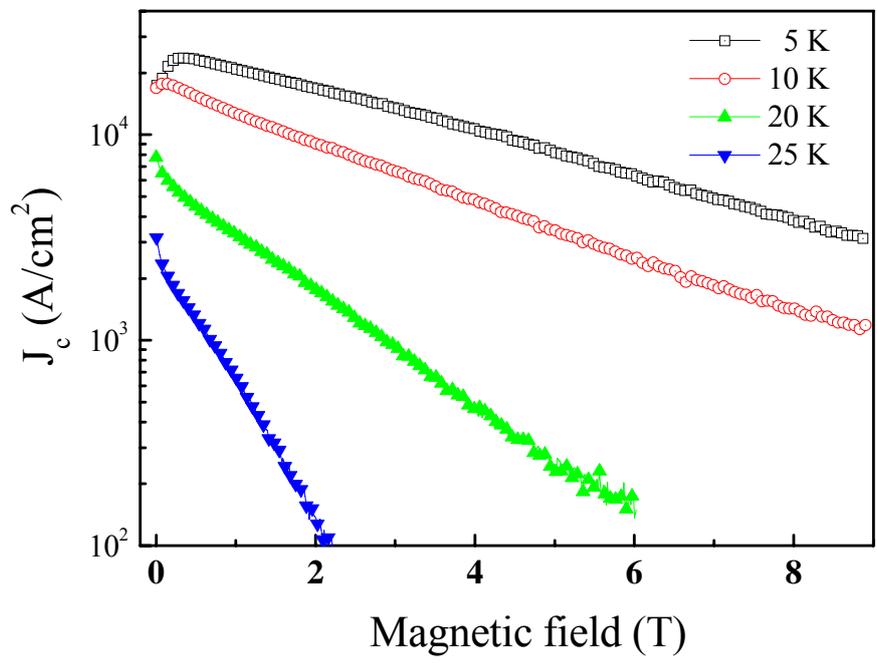

Fig.6 Gao et al.